\documentclass[11pt,a4paper]{article}
\pdfoutput=1

\usepackage{amsmath,amssymb,tikz}
\usepackage{graphicx}
 \allowdisplaybreaks
\def\be{\begin{equation}}
\def\ee{\end{equation}}
\def\ba#1\ea{\begin{align}#1\end{align}}
\def\bg#1\eg{\begin{gather}#1\end{gather}}
\def\bm#1\em{\begin{multline}#1\end{multline}}
\def\bmd#1\emd{\begin{multlined}#1\end{multlined}}

\def\a{\alpha}
\def\b{\beta}

\def\e{\epsilon}

\def\({\left(}
\def\){\right)}
\def\[{\left[}
\def\]{\right]}

\def\Tr{{\rm Tr}}

\def \be {\begin{equation}}
\def \ee {\end{equation}}
\def \ba {\begin{array}}
\def \ea {\end{array}}
\def \bea{\begin{eqnarray}}
\def \eea{\end{eqnarray}}

\def \a {\alpha}
\def \b {\beta}

\def \e {\epsilon}

\def\bea{\begin{eqnarray}}
\def\eea{\end{eqnarray}}

\newcommand{\bit}{\begin{itemize}}  \newcommand{\eit}{\end{itemize}}
\newcommand{\ben}{\begin{enumerate}}  \newcommand{\een}{\end{enumerate}}

\long\def\symbolfootnote[#1]#2{\begingroup%
\def\thefootnote{\fnsymbol{footnote}}\footnote[#1]{#2}\endgroup}

\newcommand{\sysu}{{\it School of Physics and Astronomy, Sun Yat-Sen University, 2 Daxue Road, Zhuhai 519082, China}}

\begin{document}
\thispagestyle{empty}
\begin{center}

~\vspace{20pt}

{\Large\bf Casimir Effect, Weyl Anomaly and Displacement Operator in Boundary Conformal Field Theory}

\vspace{25pt}

Rong-Xin Miao ${}$\symbolfootnote[1]{Email:~\sf
  miaorx@mail.sysu.edu.cn}

\vspace{10pt}${}$\sysu

\vspace{2cm}

\begin{abstract}

In this paper, we investigate Casimir effect, Weyl anomaly and displacement operator for boundary conformal field theory in general dimensions. We find universal relations between them. In particular, they are all determined by the central charge of boundary conformal field theory. We verify these relations by studying free BCFTs and holographic BCFTs.  As a byproduct, we obtain the holographic two point function of stress tensor when the bulk boundary is perpendicular to the AdS boundary. 

\end{abstract}

\end{center}

\newpage
\setcounter{footnote}{0}
\setcounter{page}{1}

\tableofcontents

\section{Introduction}

The boundary brings many novel characteristics to quantum field theories. The most famous one is the Casimir effect \cite{Casimir:1948dh,Plunien:1986ca,Bordag:2001qi}, which originates from the change of zero point energy of quantized fields due to boundaries. In this paper we focus on BCFT, the conformal field theory defined on a manifold $M$ with a boundary $\partial M$ and suitable boundary conditions (BC) \cite{Cardy:2004hm, McAvity:1993ue}. 
We use $x^i$ and $y^a$ to denote the coordinates of $M$ and $\partial M$, respectively. We have $x^i=(x,y^a)$ and the boundary is labeled by $x=0$.  
It is remarkable that the renormalized expectation value of stress tensor of BCFT is divergent near the boundary \cite{Deutsch:1978sc},
\begin{eqnarray}\label{Tij}
<T_{ij}>= \alpha \frac{ \bar{k}_{ij}}{x^{d-1}}+O(\frac{1}{x^{d-2}}), \ \ x\sim 0,
\end{eqnarray}
where $x$ is the distance to the boundary, $\bar{k}_{ij}$ is the traceless part of extrinsic curvature, $d$ is the dimension of spacetime and $\alpha$ is a constant determined by the type of BCFT under consideration. One may worry about the divergence of stress tensor at $x=0$. In fact, nothing goes wrong, since there are boundary contributions to the stress tensor, which exactly cancel the apparent bulk “divergence” and make finite the total energy \cite{Kennedy:1979ar, Miao:2017aba}. 
Roughly speaking, we call the renormalized stress tensor (1) as `Casimir effect' in this paper.

In addition to Casimir energy, the boundary also plays an important role in the quantum anomaly \cite{
  Fursaev:2015wpa,Herzog:2015ioa,Herzog:2017kkj,Herzog:2017xha,Miao:2017gyt,Chu:2017aab,
Jensen:2017eof, 
Kurkov:2017cdz,Kurkov:2018pjw,Vassilevich:2018aqu,Rodriguez-Gomez:2017kxf,Jensen:2015swa,Solodukhin:2015eca}. 
Let us focus on Weyl anomaly, which measures the  breaking of scaling symmetry of CFT due to quantum effects,
\begin{eqnarray}\label{definationA}
\mathcal{A}= \partial_{\sigma} I_{\text{eff}}[e^{2\sigma} g_{ij}]|_{\sigma=0}=\int_M dx^d\sqrt{g}<T^i_i>+\int_{\partial M}dy^{d-1}\sqrt{h}<t^a_a>,
\end{eqnarray}
where $I_{\text{eff}}$ is the effective action of CFT, $g_{ij}$ are the bulk metrics, $h_{ab}$ are the induced metrics on the boundary, $T_{ij}$ and $t_{ab}$ are bulk and boundary renormalized stress tensors,  respectively. 
Note that  there are non-trivial boundary contributions to Weyl anomaly for BCFT \cite{Fursaev:2015wpa,Herzog:2015ioa}. Take 3d BCFT as an example, we have \cite{
  Jensen:2015swa,Solodukhin:2015eca}
  \begin{eqnarray}\label{3dA}
\mathcal{A}_{3d}= \int_{\partial M} dy^{2}\sqrt{h} \left(\beta_0  \mathcal{R} - \beta \text{tr}(\bar{k}^2)\right), 
\end{eqnarray}
where $\beta_0$, $ \beta$ are boundary central charges and $\mathcal{R}$ is Ricci scalar on the boundary.
 As for $d \ge 4$, in general, (\ref{definationA}) takes the form
\begin{eqnarray}\label{A}
\mathcal{A}= \text{ Bulk Anomaly}+\int_{\partial M} dy^{d-1}\sqrt{h} \left( \beta \bar{k}^{ij} q^{(d-2)}_{ij}+... \right),
\end{eqnarray}
 where $\beta$ is the boundary central charge, $...$ denote terms without traceless parts of $q^{(d-2)}_{ij}$, $q^{(n)}_{ij}$ is defined by the near-boundary metric in the Gauss normal coordinate 
 \begin{eqnarray}\label{Gaussmetric}
ds^2=dx^2+(h_{ab}-2x k_{ab}+\sum_{n=2}^{\infty}x^n q^{(n)}_{ab}) dy^a dy^b,
\end{eqnarray}
where $k_{ab}=\frac{\partial x^i}{\partial y^a} \frac{\partial x^j}{\partial y^b} k_{ij}$ and similar for $q^{(n)}_{ab}$. The covariant form of Weyl anomaly of 4d BCFT can be found in \cite{Fursaev:2015wpa,Herzog:2015ioa}. In particular, we have $\bar{k}^{ij} C_{i k j l}h^{kl}=\frac{1}{2} \bar{k}^{ij} q^{(2)}_{ij}+  \text{terms without}\  \bar{q}^{(2)}_{ij}$ for $d=4$ \cite{Chu:2017aab}, where $C_{ijkl}$ are Weyl tensors.
Since it is a non-trivial problem to find out the exact expressions of boundary Weyl anomaly in general dimensions, for simplicity we focus on the non-covariant form (\ref{A}) in this paper. 

Because the boundary breaks the translation invariance along the direction perpendicular to the boundary, the energy moment tensor of BCFT is no longer conserved generally. Instead, we have \cite{Billo:2016cpy}
\begin{eqnarray}\label{DT}
\nabla_i T^{ij}=-\delta(x)D^j(y^a),
\end{eqnarray}
 where $D^j(y^a)$ is the displacement operator with scaling dimension $\Delta=d$. Note that only the orthogonal component of displacement operator is non-zero, i.e., $D^in_i=-D^x\ne 0$, where $n_i$ is the outward-pointing normal vector of the boundary. The two point function of displacement operator  is given by
 \begin{eqnarray}\label{2pointD}
<D^x(y)D^x(0)>= \frac{C_D}{|y|^{2d}},
\end{eqnarray}
with $C_D$ the Zamolodchikov norm, which is a piece of BCFT data \cite{Billo:2016cpy}.

The main goal of this paper is to show that the Casimir effect, Weyl anomaly and displacement operator of BCFT are closely related to each other. As we will prove in the text, there are universal relations
 \begin{eqnarray}\label{universalrelation1}
&&\alpha=2\beta \\
&&\alpha=-\frac{d\Gamma[\frac{d+1}{2}]\pi^{\frac{d-1}{2}}}{(d-1)\Gamma[d+2]}C_D\label{universalrelation2}.
\end{eqnarray}
between the charges of Casimir effect (\ref{Tij}), Weyl anomaly (\ref{A}) and displacement operator (\ref{2pointD}). We also give some holographic and free-field tests of the above universal relations. 

It should be stressed that the above universal relations (\ref{universalrelation1}, \ref{universalrelation2}) are generalizations of the works \cite{Miao:2017aba,Herzog:2017kkj,Herzog:2017xha} to higher dimensions. \cite{Miao:2017aba} find universal relations between Casimir coefficient $\alpha$ and boundary central charge $\beta$ for 3d and 4d BCFTs. And \cite{Herzog:2017kkj,Herzog:2017xha} find remarkable relations between  boundary central charge $\beta$ and the Zamolodchikov norm $C_D$ of displacement operator in three and four dimensions.  As we will show in the paper, the universal relations (\ref{universalrelation1}, \ref{universalrelation2}) agree with the results of \cite{Miao:2017aba,Herzog:2017kkj,Herzog:2017xha}. Besides, it should be also mentioned that our results are generalization of the works of \cite{Lewkowycz:2014jia,Bianchi:2015liz,Dong:2016wcf,Bianchi:2016xvf,Chu:2016tps,Balakrishnan:2016ttg} for codimension 2 defect (entangling surface) to codimension 1 defect (boundary).

The paper is organized as follows. 
In section 2, we briefly review the displacement operator of BCFT. In section 3, we study the universal relation between Casimir energy and Weyl anomaly. In section 4, we derive the shape dependence of Casimir effect from displacement operator. In section 5,  we verify our main results by studying free BCFTs and holographic BCFTs. Finally, we conclude with discussions in section 6.

\section{Review of displacement operator}

In this section we briefly review the displacement operator for BCFT. Consider the variation of the effective action of BCFT, in general, we have \cite{McAvity:1993ue,Billo:2016cpy}
\begin{eqnarray}\label{dID}
\delta I_{\text{eff}}=\frac{1}{2}\int_M dx^d \sqrt{g} T_{\text{bulk}}^{ij}\delta g_{ij}+\frac{1}{2}\int_{\partial M} dy^{d-1} \sqrt{h}\left( T_{\text{bdy}}^{ij}\delta h_{ij}+J^{(n)ij}\delta q^{(n)}_{ij}+2D_i \delta x^i \right)
\end{eqnarray}
where $T_{\text{bulk}}^{ij}$ and $T_{\text{bdy}}^{ij}$ are bulk and boundary stress tensor respectively, $J^{(n)ij}$ is the boundary current conjugate to $q^{(n)}_{ij}=\partial^n_x g_{ij}/\Gamma[n+1]$ (\ref{Gaussmetric}) and $D_i$ is the displacement operator. For simplicity, we turn off the variation of $ q^{(n)}_{ij}$ below. Please see \cite{McAvity:1993ue,Billo:2016cpy} for the discussions of such terms. Consider the diffeomorphism on the boundary,
\begin{eqnarray}\label{diffbdy}
\delta_{\zeta} y^a=-\zeta^a, \ \ \delta_{\zeta} x^i=\zeta^a \partial_a x^i,\ \ \delta_{\zeta} g_{ij}=0
\end{eqnarray}
we have
\begin{eqnarray}\label{dIDbdy}
\delta_{\zeta}  I_{\text{eff}}=\int_{\partial M} dy^{d-1} \sqrt{h} \zeta^a \partial_a x^i D_i=0,
\end{eqnarray}
which yields
\begin{eqnarray}\label{Dbdy}
 \zeta^a \partial_a x^i D_i=0.
\end{eqnarray}
This means the components of $D^i$ parallel to the boundary must vanish. This is the expected result, since the translation invariance along the direction parallel to the boundary is preserved. As a result, we must have $\nabla^i T_{ij}\partial_a x^j=0$, which yields (\ref{Dbdy}) from (\ref{DT}). 

Now consider an infinite-small transformation of bulk coordinates
\begin{eqnarray}\label{diffbulk}
\delta_{\xi} y^a=0, \ \ \delta_{\xi} x^i=-\xi^i,\ \  \delta_{\xi} g_{ij}=2\nabla_{(i} \xi_{j)},
\end{eqnarray}
we get 
\begin{eqnarray}\label{dIDbulk1}
\delta_{\xi}  I_{\text{eff}}&=&-\int_{ M} dx^{d} \sqrt{g} \xi_j \nabla_i T_{\text{bulk}}^{ij}\nonumber\\
&&+\int_{\partial M} dy^{d-1} \sqrt{h} \left(T_{\text{bulk}}^{ij}n_i\xi_j -\hat{\xi}_j\hat{\nabla}_iT_{\text{bdy}}^{ij} +T_{\text{bdy}}^{il}k_{il}n^j \xi_j-\xi_j D^j \right)\nonumber\\
&=&0,
\end{eqnarray}
where $\hat{\nabla}_i$ is the induced covariant derivative on the boundary and $\hat{\xi}_i=\xi_l h^l_i$ is the pull back of the bulk vector $\xi_i$ into the boundary. We have used $T_{\text{bdy}}^{ij}n_j=0$ in the above derivations. For infinite-small bulk $\xi$, we derive from (\ref{dIDbulk1})
\begin{eqnarray}\label{bulklaw}
\nabla_i T_{\text{bulk}}^{ij}=0.
\end{eqnarray}
As for infinite-small boundary $\xi$, we get 
\begin{eqnarray}\label{boundarylaw1}
&&\hat{\nabla}_iT_{\text{bdy}}^{ij}=T_{\text{bulk}}^{il}n_i h_l^j , \\
&& n_j D^j=T_{\text{bulk}}^{ij}n_in_j +T_{\text{bdy}}^{ij}k_{ij}, \label{boundarylaw2}
\end{eqnarray} 
which agrees with \cite{Jensen:2015swa}. Note that there could be corrections to (\ref{boundarylaw1},\ref{boundarylaw2}) if we turn on the variation of $ q^{(n)}_{ij}$ \cite{McAvity:1993ue}. Below we focus on the flat space with a plate boundary. i.e., $k_{ij}=q^{(n)}_{ij}=0$. Then the displacement operator $D=D_x=-n_j D^j$ becomes
\begin{eqnarray}\label{dispalcementkey}
D(y)=-T_{\text{bulk}}^{ij}n_in_j= -T_{xx}(0,y),
\end{eqnarray} 
where $T^{ij}=T_{\text{bulk}}^{ij}+\delta(x) T_{\text{bdy}}^{ij}$ is the total stress tensor. From (\ref{dispalcementkey}), it is clear that the displacement operator of BCFT is given by the normal component of the stress tensor in the flat space with a plate boundary \cite{McAvity:1993ue,Billo:2016cpy,Jensen:2015swa}.  As a result, we have
\begin{eqnarray}\label{dispalcementpoint1}
&&<D(y)>=-<T_{xx}(0,y)>=0,\nonumber\\
&&<D(y_1)D(y)>=<T_{xx}(0,y_1)T_{xx}(0,y)>=\frac{\alpha(1)}{|y_1-y|^{2d}}, \label{dispalcementpoint2}
\end{eqnarray} 
where $\alpha(1)=C_D$ is defined by (2.34) of \cite{McAvity:1993ue}. 

Now let us go on to discuss the two point functions. To  reveal the relation between Casimir effect and displacement operator, we need the correlator of displacement operator with the stress tensor. According to  \cite{Billo:2016cpy}, we have
\begin{eqnarray}\label{TD1}
&&<T^{ab}(x_1) D(y)>=b\left( \frac{4x_1^2 y^ay^b}{(x_1^2+y^2)^{d+2}}-\frac{\delta^{ab}}{d(x_1^2+y^2)^{d}}\right),\\ \label{TD2}
&&<T^{ax}(x_1) D(y)>=2b\left( \frac{ y^a x_1}{(x_1^2+y^2)^{d+1}}-\frac{2y^{a}x_1^3}{(x_1^2+y^2)^{d+2}}\right),\\
&&<T^{xx}(x_1) D(y)>= \frac{ b}{(x_1^2+y^2)^{d}}\left( \frac{(x_1^2-y^2)^2}{(x_1^2+y^2)^{2}}-\frac{1}{d}\right), \label{TD3}
\end{eqnarray} 
where $D(y)=D(x=0,y^a)$ and $T^{ij}(x_1)=T^{ij}(x_1,y_1^a=0)$. Taking the limit $x_1\to 0$ for (\ref{TD3}) and comparing with (\ref{dispalcementpoint2}), we get
\begin{eqnarray}\label{TDb}
b=-\frac{d}{(d-1)}\alpha(1)=-\frac{d}{(d-1)}C_D.
\end{eqnarray} 

\section{Casimir effect from Weyl anomaly}

In \cite{Miao:2017aba}, it is found that there are universal relations between Casimir effect and Weyl anomaly for BCFTs in three and four dimensions. In this section, we generalized the results of \cite{Miao:2017aba} to higher dimensions. 

The key observation of \cite{Miao:2017aba} is that, the Weyl anomaly $\mathcal{A}$ can be obtained  either as the trace of
renormalized stress tensor or the logarithmic
UV divergent term of the
effective action. Thus if we vary the metric and focus on the boundary term, we obtain
\begin{eqnarray}\label{dI}
 (\delta \mathcal{A})_{\partial M}  =\delta I_{\text{eff}}\big|_{\ln 1/\epsilon}=
  \frac{1}{2}\int_M \sqrt{g}T^{ij}\delta g_{ij}\big|_{\ln 1/\epsilon},
\end{eqnarray}
where $\e$ is an UV cutoff.  To proceed, let us focus on the metric in the Gauss normal coordinates (\ref{Gaussmetric}). For simplicity, we only turn on the variation with respect to $q^{(d-2)}_{ab}$, i.e., $\delta g_{ab}= x^{d-2} \delta q^{(d-2)}_{ab}, \delta g_{xx}=\delta g_{xb}=0$.

Let us firstly discuss the case $d\ge 4$, where $q^{(d-2)}_{ab}$ and $k_{ab}$ are independent. 
From (\ref{A}), we derive the left hand side of (\ref{dI}) as
\begin{eqnarray}\label{dIleft}
(\delta \mathcal{A})_{\partial M} =\beta \int_{\partial M} dy^{d-1}\sqrt{h} \bar{k}^{ab} \delta q^{(d-2)}_{ab}.
\end{eqnarray}
From (\ref{Tij}) together with $\delta g_{ab}= x^{d-2} \delta q^{(d-2)}_{ab}$, we obtain the right hand side of (\ref{dI}) as
\begin{eqnarray}\label{dIright}
\delta I_{\text{eff}}\big|_{\ln 1/\epsilon}&=&
  \frac{\alpha}{2}\int_M dx dy^{d-1}\sqrt{h} \frac{ \bar{k}^{ab}}{x}\delta q^{(d-2)}_{ab}\big|_{\ln 1/\epsilon}\nonumber\\
  &=&\frac{\alpha}{2} \int_{\partial M} dy^{d-1}\sqrt{h} \bar{k}^{ab} \delta q^{(d-2)}_{ab}.
\end{eqnarray}
Identifying (\ref{dIleft}) with (\ref{dIright}), we obtain the universal relation (\ref{universalrelation1}) for $d\ge 4$.

Now let us go on to discuss the case $d=3$, where $q^{(d-2)}_{ab}=q^{(1)}_{ab}=-2 k_{ab}$ and $k_{ab}$ are not independent. From (\ref{3dA}), we get the left hand side of (\ref{dI}) as
\begin{eqnarray}\label{dIleft}
(\delta \mathcal{A})_{\partial M} =-2\beta \int_{\partial M} dy^{2}\sqrt{h} \bar{k}^{ab} \delta k_{ab}.
\end{eqnarray}
Note that $\int_{\partial_M} dx^2\sqrt{h}\mathcal{R}$ in (\ref{3dA}) is the Euler density, whose variation vanishes.
From (\ref{Tij}) together with $\delta g_{ab}= -2x \delta k_{ab}$, we derive the right hand side of (\ref{dI}) as
\begin{eqnarray}\label{dIright}
\delta I_{\text{eff}}\big|_{\ln 1/\epsilon}&=&
  -\alpha\int_M dx dy^{2}\sqrt{h} \frac{ \bar{k}^{ab}}{x}\delta k_{ab}\big|_{\ln 1/\epsilon}\nonumber\\
  &=&-\alpha \int_{\partial M} dy^{2}\sqrt{h} \bar{k}^{ab} \delta k_{ab}.
\end{eqnarray}
Identifying (\ref{dIleft}) with (\ref{dIright}), we obtain  (\ref{universalrelation1}) for $d=3$. Now we finish the derivations of the universal relation (\ref{universalrelation1}) between Casimir effect and Weyl anomaly in general dimensions. 

Now we show that the universal relation (\ref{universalrelation1})  agree with the results of \cite{Miao:2017aba} in three and four dimensions. In the notations of  \cite{Miao:2017aba}, the universal laws between Casimir effect and Weyl anomaly are given by $\alpha_1=-b_2$ for 3d BCFTs and $\alpha_1=b_4/2$ for 4d BCFTs.  Transforming into our notations, i.e., $\alpha_1=\alpha/2, b_2=-\beta, b_4=2\beta$, the universal laws of \cite{Miao:2017aba} become $\alpha=2\beta$ for both 3d and 4d BCFTs, which is exactly the universal relation  (\ref{universalrelation1})  in our notations.  Note that, to transform the notation $b_4=2\beta$, we have used $\bar{k}^{ij} C_{i k j l}h^{kl}=\frac{1}{2} \bar{k}^{ij} q^{(2)}_{ij}+  \text{terms without}\  \bar{q}^{(2)}_{ij}$ \cite{Chu:2017aab}.

\section{Casimir effect from displacement operator}

In this section, we derive the shape dependence of Casimir effect from the displacement operator. 
Note that the technology used in this section was used in  \cite{Bianchi:2016xvf}  to determine the relation between $C_D$ and the one-point function of the stress tensor in the presence of a defect of codimension two.

By definition, we have for displacement operator \cite{Bianchi:2016xvf} 
\begin{eqnarray}\label{Dpoints}
<D ...>=n_i \frac{\delta}{\delta x^i}<...>,
\end{eqnarray}
where $...$ denote arbitrary insertions of operators. From (\ref{Dpoints}), one can derive the one point function of an operator near the deformed boundary from the two point function of this operator with displacement operator on the non-deformed boundary. Take stress tensor as an example, we have
\begin{eqnarray}\label{keyformulaD}
<T_{ij}(x_1)>_{f\partial M}=<T_{ij}(x_1)>_{\partial M}-\int dy^{d-1}<T_{ij}(x_1)D(y)>_{\partial M} f(y)+O(f^2),
\end{eqnarray}
where we have $\delta x^i=\delta^i_x f(y)$ and recall that we have set $y_1^a=0$ for $T_{ij}$ and $x=0$ for $D$ for simplicity. For a general deformation the above integral cannot be performed. Instead, only the singular parts near the boundary $x_1\to 0$ can be calculated explicitly. In the weak sense, i.e., after integration against a test function, the correlator (\ref{TD1}) can be rewritten as distributions with support on the boundary. In the appendix, we prove
\begin{eqnarray}\label{TDW1}
&&<T_{ab}(x_1) D(y)>=\frac{b\Gamma[\frac{d+1}{2}]\pi^{\frac{d-1}{2}}}{\Gamma[d+2]}\left( \frac{\partial_a\partial_b \delta^{d-1}(y)-\delta_{ab}\frac{1}{d-1}\partial^2 \delta^{d-1}(y)}{(x_1)^{d-1}}+...  \right),\\ \label{TDW2}
&&<T_{ax}(x_1) D(y)>=-\frac{b\Gamma[\frac{d+1}{2}]\pi^{\frac{d-1}{2}}}{(d-1)\Gamma[d+2]}\left( \frac{\partial_a\partial^2 \delta^{d-1}(y)}{x_1^{d-2}}+...  \right),\\
&&<T_{xx}(x_1) D(y)>=\left( 0+...\right), \label{TDW3}
\end{eqnarray} 
where $...$ denote higher order terms and the terms without derivatives of delta function (we focus on the case $\delta^{d-1}(y) f(y)=0$ below). 
Substituting (\ref{TDW1},\ref{TDW2},\ref{TDW3}) into (\ref{keyformulaD}) and using $\partial_a\partial_bf(y)=-k_{ab}(y)$ together with $<T_{ij}>_{\partial M}=0$, we obtain
\begin{eqnarray}\label{keyresult1}
&&<T_{ab}(x)>_{f\partial M}=\frac{b\Gamma[\frac{d+1}{2}]\pi^{\frac{d-1}{2}}}{\Gamma[d+2]}\ \frac{\bar{k}_{ab}}{x^{d-1}}+O(k^2),\\ \label{keyresult2}
&&<T_{ax}(x)>_{f\partial M}=\frac{b\Gamma[\frac{d+1}{2}]\pi^{\frac{d-1}{2}}}{(d-1)\Gamma[d+2]}\ \frac{\partial_a\bar{k}}{x^{d-2}}+O(k^2),\\
&&<T_{xx}(x)>_{f\partial M}=O(k^2),\label{keyresult3}
\end{eqnarray}
where we have replaced $x_1$ by $x$ for simplicity. 
These are some of the main results of this paper. Comparing (\ref{keyresult1}) with (\ref{Tij}) and recalling that $b=-\frac{d}{(d-1)}C_D$ (\ref{TDb}), we derive the universal relation (\ref{universalrelation2}) between Casimir effect and displacement operator. Note that we only list $T_{ij}$ up to order $O(1/ x^{d-1})$ in (\ref{Tij}). To the next order in flat space, we have \cite{Deutsch:1978sc,Miao:2017aba}
\begin{eqnarray}\label{Tij2}
&&T_{ax}=\frac{\a}{d-1} \ \frac{\partial_a\bar{k}}{x^{d-2}}+O(\frac{1}{x^{d-3}}), \label{SolutionTij2}\\
&&T_{xx}=\frac{\a}{d-2} \ \frac{\Tr \bar{k}^2 }{x_1^{d-2}} +O(\frac{1}{x^{d-3}}) \label{SolutionTij2},\label{Tij3}
\end{eqnarray} 
which agree with (\ref{keyresult2},\ref{keyresult3}) and (\ref{universalrelation2}). This can be regarded as a double-check of our calculations. 
It should be mentioned that, for 3d and 4d BCFTs, \cite{Herzog:2017kkj,Herzog:2017xha} find interesting relation between Weyl anomaly and displacement operator, while \cite{Miao:2017aba} obtain universal relation between Weyl anomaly and Casimir effect. Combining their results, we can verify our main result (\ref{universalrelation2}) between displacement operator and Casimir effect for $d=3,4$.  
Let us show more details below. Since we have already shown that results of \cite{Miao:2017aba} agree with ours at the end of sect. 3, now we focus on the results of \cite{Herzog:2017kkj,Herzog:2017xha}. In the notations of  \cite{Herzog:2017kkj,Herzog:2017xha}, the universal laws between Weyl anomaly and displacement operator are expressed as
\begin{eqnarray}\label{compare1112a}
&& b=\frac{\pi^2}{8} c_{nn},\  \ \ \ \text{ for d=3}, \\
&& b_2=\frac{2\pi^4}{15} c_{nn},\  \text{ for d=4},\label{compare1112b}
\end{eqnarray} 
where $b, b_2$ are boundary central charges and $c_{nn}$ denotes the norm of displacement operator.
Transforming into our notations, i.e., $b=-4\pi \beta, b_2=-32\pi^2 \beta, c_{nn}=C_D$, (\ref{compare1112a},\ref{compare1112b}) become
\begin{eqnarray}\label{compare1112A}
&& \beta=-\frac{\pi}{32} C_D, \ \ \text{ for d=3}, \\
&& \beta=-\frac{\pi^2}{240} C_D,\  \text{ for d=4}.\label{compare1112B}
\end{eqnarray} 
 Note that, to transform the notation $ b_2=-32\pi^2 \beta$, we have used $\bar{k}^{ij} C_{i k j l}h^{kl}=\frac{1}{2} \bar{k}^{ij} q^{(2)}_{ij}+  \text{terms without}\  \bar{q}^{(2)}_{ij}$ \cite{Chu:2017aab}. From our key results (\ref{universalrelation1},\ref{universalrelation2}), we get
  \begin{eqnarray}\label{comparemine}
\beta=-\frac{d\Gamma[\frac{d+1}{2}]\pi^{2\frac{d-1}{2}}}{(d-1)\Gamma[d+2]}C_D,
\end{eqnarray}
 which reduces to (\ref{compare1112A}) and (\ref{compare1112B}) for 3d BCFT and 4d BCFT,  respectively.  Now we have shown that our results indeed agree with those of \cite{Herzog:2017kkj,Herzog:2017xha} in three and four dimensions. This is a test of our universal results in general dimensions.

\section{Tests of universal relations}

\subsection{Story of free BCFT}

Now let us verify our results by studying free BCFT. For simplicity, we focus on conformally coupled free scalar with the following action 
\begin{eqnarray}\label{scalaraction}
I=\frac{1}{2}\int_{M}dx^d\sqrt{g} (\nabla_i \phi \nabla^i \phi+\xi R \phi^2)+\int_{\partial M} dy^{d-1}\sqrt{h} \xi k \phi^2,
\end{eqnarray} 
where $\xi=\frac{d-2}{4(d-1)}$ and $k$ is the extrinsic curvature. There are two kinds of conformally invariant BCs for free scalar
\begin{equation}\label{BCscalar}
\begin{split}
&\text{Dirichlet BC} : \phi|_{\partial M}=0,\\
&\text{Robin BC} : \ \ (\nabla_n + 2\xi k)\phi|_{\partial M}=0.
\end{split}
\end{equation}
We use the heat kernel \cite {Vassilevich:2003xt} to derive the renormalized stress tensor. The heat kernel of scalar satisfies the EOM
 \begin{eqnarray}\label{EOM}
\partial_t K(t,x_i,x'_i)-(\Box-\xi R) K(t,x_i,x'_i)=0
\end{eqnarray}
 together with the BC (\ref{BCscalar}) at $x=0$ and BC
 \begin{eqnarray}\label{BCt}
\lim_{t\to 0} K(t,x_i,x'_i)=\delta^{d}(x_i-x'_i)
\end{eqnarray}
at $t=0$.  Using the heat kernel, we can obtain the Green function
 \begin{eqnarray}\label{Greenfunction}
G(x_i,x'_i)=\int_0^{\infty} dt K(t,x_i,x'_i),
\end{eqnarray}
and then derive the expectation value of the stress tensor by
\begin{eqnarray}\label{hatTij}
\hat{T}_{ij}=\lim_{x'_i\to x_i} \left[  (1-2\xi) \nabla_{i}\nabla_{j'}-2\xi \nabla_{i} \nabla_{j} +(2\xi -\frac{1}{2})g_{ij} \nabla_{l} \nabla^{l'}  +\xi (R_{ij}+\frac{4\xi-1}{2}Rg_{ij}) \right] G(x_i,x'_i).\nonumber\\
\end{eqnarray}
In general $\hat{T}_{ij}$ is divergent, which can be renormalized by subtracting the value it would have in the space without boundary,
\begin{eqnarray}\label{renTij}
T_{ij}=\hat{T}_{ij}-\hat{T}_{0ij}.
\end{eqnarray}

To proceed,  we choose the following background metric
\begin{eqnarray}\label{scalarmetric}
ds^2=dx^2+(1-2k x)dy_1^2+dy_2^2+...+dy_{d-1}^2
\end{eqnarray} 
with $k$ a constant. Then we solve the heat kernel (\ref{EOM},\ref{BCscalar},\ref{BCt}) perturbatively in powers of $k$. At the linear order, we get 
\begin{eqnarray}\label{heatkernel}
K(t,x,x')=\frac{1}{(4\pi t)^{\frac{d}{2}}}\left(\exp[-\frac{\rho}{4t}]+ \exp[-\frac{\rho_I}{4t}]\Omega_I (t) \right) + O(k^2),
\end{eqnarray} 
where $\rho$ and $\rho_I$ are the geodesic distances in real space and image space, respectively,
\begin{eqnarray}\label{distance1}
&&\rho=(x-x')^2+(y_a-y'_a)^2-k (x'+x) \left(y_1-y_1'\right){}^2,\\
&&\rho_I=(x+x')^2+(y_a-y'_a)^2-k\frac{(x^2+x'^2) \left(y_1-y_1'\right){}^2}{x'+x} \label{distance2}.
\end{eqnarray} 
For Dirichlet BC, $\Omega_I (t)$ is given by
\begin{eqnarray}\label{omegaDBC}
&&\Omega_I=-1\nonumber\\
&&+k \frac{x x'}{4 t}\left(\left(y_1-y_1'\right){}^2 (\frac{\sqrt{\pi } e^{\frac{\left(x'+x\right)^2}{4 t}} \text{erfc}\left(\frac{x'+x}{2 \sqrt{t}}\right)}{\sqrt{t}}-\frac{2}{x'+x})-2 \sqrt{\pi } \sqrt{t} e^{\frac{\left(x'+x\right)^2}{4 t}} \text{erfc}\left(\frac{x'+x}{2 \sqrt{t}}\right)\right).\nonumber\\
\end{eqnarray} 
As for Robin BC, $\Omega_I (t)$ reads
\begin{eqnarray}\label{omegaRBC}
&&\Omega_I=1\nonumber\\
&&+k\big[ \left(y_1-y_1'\right){}^2 (\frac{\sqrt{\pi } e^{\frac{\left(x'+x\right)^2}{4 t}} \text{erfc}\left(\frac{x'+x}{2 \sqrt{t}}\right)}{\sqrt{t}}-\frac{2}{x'+x})-2 \sqrt{\pi } \sqrt{t} e^{\frac{\left(x'+x\right)^2}{4 t}} \text{erfc}\left(\frac{x'+x}{2 \sqrt{t}}\right) \nonumber\\
&& \left(y_1-y'_1\right){}^2 (\frac{\sqrt{\pi } e^{\frac{\left(x+x'\right){}^2}{4 t}} \left(2 t+x^2+x'^2\right) \text{erfc}\left(\frac{x+x'}{2 \sqrt{t}}\right)}{8 t^{3/2}}+\frac{x x'}{2 t \left(x+x'\right)}-\frac{x+x'}{4 t}) \big].
\end{eqnarray} 
 Note that the above heat kernel (\ref{heatkernel}) agrees with the general results of \cite{McAvity:1990we}. We remark that the first term of (\ref{heatkernel}) is the heat kernel in free space without boundary, while the second term of (\ref{heatkernel}) is due to the boundary effect. To calculate the renormalized stress tensor (\ref{renTij}), we subtract the first term and only keep the second term of (\ref{heatkernel}). 

Substituting the heat kernel (\ref{heatkernel},\ref{distance2},\ref{omegaDBC},\ref{omegaRBC}) into (\ref{Greenfunction},\ref{hatTij},\ref{renTij}), after some complicated calculations, we obtain
the renormalized stress tensor
\begin{eqnarray}\label{renTijscalar}
T_{ij}=\frac{2^{-d} \pi ^{-\frac{d}{2}} \Gamma \left(\frac{d}{2}\right)}{1-d^2}\frac{\bar{k}_{ij}}{x^{d-1}}+O(k^2)
\end{eqnarray}
which gives
\begin{eqnarray}\label{alphascalar}
\alpha=\frac{2^{-d} \pi ^{-\frac{d}{2}} \Gamma \left(\frac{d}{2}\right)}{1-d^2}
\end{eqnarray}
for free scalar. It is remarkable that Dirichlet BC and Robin BC yield the same $\alpha$. Actually, this is a special character of free BCFT. In general $\alpha$ depends on BCs \cite{Miao:2017aba,Miao:2018qkc}.

$C_D$ of free scalar is calculated in \cite{McAvity:1993ue}, which is given by
\begin{eqnarray}\label{CDscalar}
C_D=\alpha(1)=\frac{\Gamma[\frac{d}{2}]^2}{2\pi^d}. 
\end{eqnarray}
One can check that (\ref{alphascalar}) and (\ref{CDscalar}) indeed satisfy the universal relation (\ref{universalrelation2}) between Casimir effect and displacement operator.  The universal relation (\ref{universalrelation1}) between Casimir effect and Weyl anomaly has been verified for 3d and 4d free BCFT in \cite{Miao:2017aba}. Since Weyl anomaly of free BCFT in higher dimensions is unknown in the literature, so far we cannot verify the universal relation (\ref{universalrelation1}) generally.  In the next subsection, we shall test  (\ref{universalrelation1}) by studying holographic BCFTs.

\subsection{Holographic BCFT}

The bottom-up model of holographic BCFT is firstly studied by Takayanagi \cite{Takayanagi:2011zk}. Neumann boundary condition (NBC) plays an important role in this model and produces many interesting results \cite{Nozaki:2012qd,Fujita:2011fp}. 
In this section, we use the holographic model of BCFT  \cite{Takayanagi:2011zk} to test the universal relations  (\ref{universalrelation1}, \ref{universalrelation2}).

\begin{figure}[t]
\centering
\includegraphics[width=5cm]{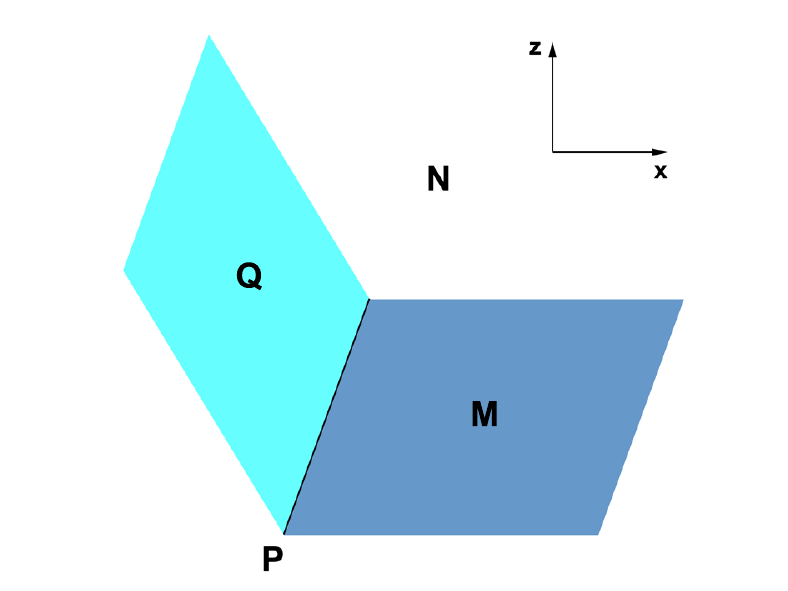}
\caption{Geometry of holographic BCFT}
\label{MNPQ}
\end{figure}

Let us start with the geometry setup of holographic BCFT. The $d$ dimensional manifold $M$ is extended to a $d+1$
dimensional asymptotically AdS space $N$ so that $\partial N= M\cup
Q$, where $Q$ is a $d$ dimensional manifold which satisfies $\partial
Q=\partial M=P$. See figure \ref{MNPQ} for example. A central issue in the construction of the
AdS/BCFT is the determination of the location of Q in the bulk.  It turns out that the location of Q can be fixed by boundary conditions (BC).

The action for holographic BCFT is given by ($16\pi G_N=1, L=1$)
\begin{eqnarray}\label{action}
  I=\int_N \sqrt{G} \Big(
  R-2 \Lambda  \Big)
  +2\int_Q \sqrt{\gamma} (K-T),
\end{eqnarray}
where $\Lambda=-\frac{d(d-1)}{2L^2}$ is the cosmological constant, $L$ is the AdS radius,  $K$ is the extrinsic curvature on $Q$ and
$T=(d-2) \tanh \rho$ is a constant parameter which can be regarded as the holographic dual
of boundary conditions of BCFT.  For simplicity, we set AdS radius $L=1$ in this paper. 
Following  \cite{Takayanagi:2011zk}, we impose NBC on the bulk boundary $Q$
\begin{eqnarray}
K_{ij}-(K-T)\gamma_{ij} =0. \label{NBC}
\end{eqnarray}
One can easily check that Poincare AdS
\begin{equation}\label{AdSmetric}
ds^2=\frac{dz^2+dx^2+\delta_{ab}dy^ady^b}{z^2},
\end{equation}
is a solution to the NBC (\ref{NBC}), provided that the embedding function of $Q$ is given by
\begin{equation}\label{Q}
x=-\sinh \rho\ z.
\end{equation}
Recall that we have $T=(d-2) \tanh \rho$.

The holographic one-point function of stress tensor is derived in \cite{Miao:2018qkc}, which takes the form of (\ref{Tij}) with $\alpha$ given by
\begin{eqnarray}
&&\alpha=\frac{2d \cosh ^d\rho}{(-\coth\rho)^d \, _2F_1\left(\frac{d-1}{2},\frac{d}{2};\frac{d+2}{2};-\text{csch}^2\rho \right)+d \cosh ^2\rho \coth\rho}\label{aN}.\\
\end{eqnarray}
 It should be mentioned that 
suitable analytic
continuation of the hypergeometric function should be taken
in order to get smooth function
at $\rho=0$. For example, we have for $d=4$, 
\begin{eqnarray}\label{a4ND}
 \alpha_{ 4}=\frac{-1}{(1+\tanh \rho) }.
\end{eqnarray}
In the following subsections, we will derive holographic Weyl anomaly and holographic displacement operator to verify the universal relations  (\ref{universalrelation1}, \ref{universalrelation2}).

\subsubsection{Holographic Weyl anomaly}

We follow the approach of \cite{Miao:2018qkc,Chu:2018ntx} to derive the holographic Weyl anomaly \cite{Henningson:1998gx} for BCFTs. For our purpose, we only need to work out the linear terms of $O(k)$ and $ O(q^{(d-2)})$ in the perturbation solutions. 
For simplicity, we take the following ansatz of metric
\begin{eqnarray}\label{bulkmetric}
&& ds^2=\frac{1}{z^2}\Big{[} dz^2+ \left(1+x^{d-1} \bar{k}^{ab}\bar{q}^{(d-2)}_{ab}
  X(\frac{z}{x})+...\right)dx^2  \nonumber \\
&& +\Big{(}\delta_{ab}-2x \bar{k}_{ab} f_1(\frac{z}{x}) + x^{d-2} \bar{q}^{(d-2)}_{ab} f_2(\frac{z}{x}) \nonumber\\
&& +x^{d-1} [\bar{k}_{c(a}\bar{q}^{(d-2)}{}^c_{b)}f_3(\frac{z}{x})+\delta_{ab}  \bar{k}^{ce}\bar{q}^{(d-2)}_{ce} f_4({\frac{z}{x}})]+...\Big)dy^a dy^b\Big{]}  \nonumber\\
\end{eqnarray}
where $\bar{A}_{ab}$ denote the traceless part of $A_{ab}$ and we set
\be \label{fXQ}
f_1(0)=f_2(0)=1,\quad X(0)=f_3(0)=f_4(0)=0
\ee
so
that the metric of BCFT takes the form in Gauss normal coordinates
\be \label{GNC}
ds_M^2=dx^2+ \left(\delta_{ab}-2x \bar{k}_{ab}+x^{d-2}\bar{q}^{(d-2)}_{ab}+x^{d-1} 0+...\right) dy^ady^b.
\ee
For simplicity, we focus on the solutions without $y_a$ dependence. We further
set $k_{ab}=\text{diag}(k_1,-k_1, 0,...,0),   q^{(d-2)}_{ab}=\text{diag}(q_1,-q_1, 0,...,0)$,
where $k_1, q_1$ are constants. Then the embedding function of $Q$ takes the form
\be \label{Qemb}
x=-\sinh\rho \ z + \lambda_d \bar{k}^{ab}\bar{q}^{(d-2)}_{ab} z^{d}+ ...
\ee
where $\lambda_d$ is a constant to be determined. 

 Substituting (\ref{bulkmetric}) into the Einstein equations, we get one independent equation at order $O(k)$
 \be \label{EOMf1}
s(s^2+1)f_1''(s)-(d-1)f_1'(s)=0
\ee
and another independent equation at order $O(q^{(d-2)})$
\be \label{EOMf2}
s(s^2+1)f_2''(s)-\left((d-1)+2(d-3) s^2 \right)f_2'(s)+(d-2)(d-3)s f_2(s)=0.
\ee
Solving the above equations, we obtain
 \begin{eqnarray} \label{solf1}
&&f_1(s)=1+c_1\frac{ s^d \, _2F_1\left(\frac{d-1}{2},\frac{d}{2};\frac{d+2}{2};-s^2\right)}{d},\\
&&f_2(s)=\left(s^2+1\right)^{\frac{d-3}{2}}+c_2\frac{ s^d \left(d-(d-1) \, _2F_1\left(\frac{1}{2},1;\frac{d+2}{2};-s^2\right)\right)}{d \left(s^2+1\right)}  .\label{solf2}
\end{eqnarray}
Imposing the NBC (\ref{NBC}) on $Q$ (\ref{Qemb}), we can determine the integral constants
\begin{eqnarray}
&&c_{1}=\frac{-d \cosh ^d\rho}{(-\coth\rho)^d \, _2F_1\left(\frac{d-1}{2},\frac{d}{2};\frac{d+2}{2};-\text{csch}^2\rho \right)+d \cosh ^2\rho \coth\rho},\label{c1N}\\
\end{eqnarray}
and 
\begin{eqnarray}
&&c_{2}=\frac{d \sinh^4(\rho ) \tanh (\rho ) (-\coth (\rho ))^d/(d-1)}{ (2-d+2 \cosh (2 \rho )) \, _2F_1\left(\frac{1}{2},1;\frac{d+2}{2};-\text{csch}^2\rho\right)-\frac{2 \coth ^2(\rho ) \, _2F_1\left(\frac{3}{2},2;\frac{d+4}{2};-\text{csch}^2\rho\right)}{d+2}+\frac{d (d-2-2 \cosh (2 \rho ))}{d-1} }\nonumber\\ \label{c2N}
\end{eqnarray}

Similarly, one can work out solutions of order $O(k q^{(d-2)})$. Since the solutions are quite complicated, below we focus on the case $d=4$. The generalization to higher dimensions is straightforward. It is interesting that, the integral constant $c_1$ is the same as $c_2$ for $d=4$
\begin{eqnarray}
c_{1}=c_{2}=\frac{1}{2+2 \tanh (\rho )}.
\end{eqnarray}
Solving Einstein equations of order $O(k q^{(2)})$, we get
\begin{eqnarray}
X(s)&=&\frac{1}{3} \left(\left(3 c_1 \left(2 c_1-5\right)+10\right) s^2+2 \left(1-2 c_1\right){}^2\right)+\left(1-2 c_1\right) c_1 \left(s^2+1\right) \log \left(s^2+1\right)\nonumber\\
&+&\frac{2}{3} \sqrt{s^2+1} \left(2 c_1^2 \left(s^2-2\right)+4 c_1 \left(s^2+1\right)-s^2-1\right),\label{XX}\\
f_3(s)&=&2+3 s^2-2 \sqrt{s^2+1}+c_3 \left((\sqrt{s^2+1}-\frac{3}{2}) s^2+\sqrt{s^2+1}-1\right)\nonumber\\
&+&c_1 \left(-4 s^2+8 \sqrt{s^2+1}-8\right)+c_1^2 \left(5 s^2-12 \sqrt{s^2+1}-\frac{2}{\sqrt{s^2+1}}+14\right),\label{f3}\\
f_4(s)&=&\frac{1}{9} \left(\left(\sqrt{s^2+1}-9\right) s^2+\sqrt{s^2+1}-1\right)+\frac{1}{18} c_3 \left(\left(9-6 \sqrt{s^2+1}\right) s^2-6 \sqrt{s^2+1}+6\right)\nonumber\\
&+&\frac{1}{18} c_1 \left(\left(9-8 \sqrt{s^2+1}\right) s^2-8 \sqrt{s^2+1}+3 \log \left(s^2+1\right)+8\right)\nonumber\\
&+&\frac{1}{9} c_1^2 \left(4 \sqrt{s^2+1} s^2-14 \sqrt{s^2+1}+3 \log \left(s^2+1\right)+14\right),\label{f4}
\end{eqnarray}
where we have used $c_2=c_1$ for $d=4$.  Imposing now BCs on $Q$ (\ref{Qemb}), we can fix the integral constants for NBC (\ref{NBC})
\begin{eqnarray} \label{c3N}
&&c_{3}=\frac{1}{4} e^{-2 \rho } (8 \sinh (2 \rho )+7 \cosh (2 \rho )-1),\\
&&\lambda_{4}=-\frac{-208 \sinh (2 \rho )-144 \cosh (2 \rho )+32 \cosh (4 \rho )+16 \cosh (6 \rho )+96}{1536 (\sinh (\rho )+\cosh (\rho ))^2}. \label{L4N}\nonumber\\
&&\ \ \ \ \ \ \ \ -\frac{20 \sinh (4 \rho )+16 \sinh (6 \rho )-9 \sinh (2 \rho ) \log \left(\coth ^2(\rho )\right)+3 \sinh (6 \rho ) \log \left(\coth ^2(\rho )\right)}{1536 (\sinh (\rho )+\cosh (\rho ))^2},\nonumber\\
\end{eqnarray}
where $\lambda_4$ characterizes the location of $Q$ (\ref{Qemb}).

Now we are ready to derive the holographic Weyl anomaly for 4d BCFT. 
 On-shell, the gravitational action (\ref{action}) becomes
\begin{eqnarray}\label{action5dapp}
  I=-8\int_N \sqrt{G}
  +2\int_Q \sqrt{\gamma} (K-3 \tanh \rho).
\end{eqnarray}
To get the holographic Weyl anomaly, we need to do the integration
along $x$ and $z$, and then select the UV logarithmic divergent terms.
We divide the integration region into two parts: region I is defined
by $( z \ge 0, x \ge 0)$ and region II is the complement of
region I. Let us first do the integral in region I, where only
the bulk action in (\ref{action5dapp}) contributes. Integrating along
$z$ and selecting the $1/x$ term, we obtain 
\begin{eqnarray}\label{actionregionINBC}
  I_{1}&=&-\int_{\epsilon} dx [\frac{2+\tanh\rho}{4x(1+\tanh\rho)^2} \text{Tr}(\bar{k}q^{(2)})+...]\nonumber\\
&=&-\log(\frac{1}{\epsilon})\frac{2+\tanh\rho}{4(1+\tanh\rho)^2} \text{Tr}(\bar{k}q^{(2)}) + \cdots .
\end{eqnarray}

Next let us consider the integration in region II. In this case, both the bulk
action and boundary action in (\ref{action5dapp}) contribute. For the
bulk action, we first do the integral along $x$, which yields a
boundary term on $Q$. Note that since only the UV logarithmic
divergent terms are related to Weyl anomaly, we keep only the lower
limit of the integral of $x$. Adding the boundary term from bulk
integral to the boundary action in (\ref{action5dapp}), we obtain
\begin{eqnarray}\label{actionregionIINBC}
  I_{2}&=& \int_{\epsilon} dz [\frac{\sinh(2\rho)\left(\sinh(2\rho)-\cosh(2\rho)\right)}{8z}\text{Tr}(\bar{k}q^{(2)})+...]\nonumber\\
&=&\log(\frac{1}{\epsilon}) \frac{\sinh(2\rho)\left(\sinh(2\rho)-\cosh(2\rho)\right)}{8}\text{Tr}(\bar{k}q^{(2)}) + \cdots .
\end{eqnarray} 
Adding (\ref{actionregionINBC}) to (\ref{actionregionIINBC}), we finally obtain the Weyl anomaly (\ref{A}) for 4d BCFT with the boundary central charges given by
\begin{eqnarray}\label{b4ND}
\beta_{4} =\frac{-1}{2(1+\tanh\rho)} .
\end{eqnarray}
 Comparing the above central charges with (\ref{a4ND}), we find that the universal relation (\ref{universalrelation1}) is indeed satisfied for $d=4$. It is  is straightforward to generalize the above results to higher dimensions. Following the above approach, we verify the universal relation (\ref{universalrelation1}) up to $d=6$.

\subsubsection{Holographic displacement operator}

In this section, we study the holographic two point function of displacement operator, which is equivalent to the two point function of stress tensor. That is because the displacement operator is given by the normal component of the stress tensor for BCFTs. For simplicity, we focus on the case that the bulk boundary $Q$ is perpendicular to AdS boundary $M$, i.e., $T=\rho=0$. The case with $T\ne 0$ is a non-trivial problem and we leave it to future study. 

We follow the work of \cite{Liu:1998bu} to derive the two-point function of stress tensor. Consider the metric fluctuations $H_{\mu\nu}$ in AdS spacetime
\begin{eqnarray}\label{metricH}
ds^2=\frac{dz^2+dx^2+\delta_{ab}dy^ady^b+H_{\mu\nu}dx^{\mu}dx^{\nu}}{z^2}
\end{eqnarray}
and choose the gauge
\begin{eqnarray}\label{gaugeH}
H_{zz}(z=0,{\bf{x}})=H_{zi}(z=0,{\bf{x}})=0
\end{eqnarray} 
at the AdS boundary $M$. Here the Greek letter $\mu$ denote $(z,x,y_a)$ and the Latin letter $i$ denote $(x,y_a)$.  
Imposing BCs (\ref{NBC}) on $Q$ together with the following BC on $M$
\begin{eqnarray}\label{BCH}
H_{ij}(z=0,{\bf{x}})=\hat{H}_{ij}({\bf{x}}),
\end{eqnarray} 
we solve the bulk solution
\begin{eqnarray}\label{bulkH}
H_{\mu\nu}(z,{\bf{x}})=\frac{\Gamma[d](d+1)}{\pi^{d/2}\Gamma[d/2](d-1)}\int d^dx' \Big{[}  \frac{z^d}{S^{2d}}J_{\mu i}J_{\nu j}P_{ijkl}\hat{H}_{kl}({\bf{x'}}) +  \frac{z^d}{\bar{S}^{2d}}\bar{J}_{\mu i}\bar{J}_{\nu j}P_{ijkl}\hat{H}_{kl}({\bf{x'}}) \Big]
\end{eqnarray} 
where
\begin{eqnarray}\label{fbf}
&&S^2=z^2+(x-x')^2+(y_a-y'_a)^2, \nonumber\\
&&\bar{S}^2=z^2+(x+x')^2+(y_a-y'_a)^2, \nonumber\\
&&P_{ijkl}=\frac{1}{2}\left( \delta_{ik} \delta_{jl} +\delta_{il} \delta_{jk}\right)-\frac{1}{d}\delta_{ij}\delta_{kl},\nonumber\\
&&J_{\mu\nu}=\delta_{\mu\nu}-2\frac{(x_{\mu}-x'_{\mu})(x_{\nu}-x'_{\nu})}{S^2},\nonumber\\
&&\bar{J}_{\mu\nu}=J_{\mu\nu}-2 X_{\mu} X'_{\nu},
\end{eqnarray} 
and 
\begin{eqnarray}\label{Xi}
&&X_{\mu}=\frac{1}{S\bar{S}}\left(2x z, x^2-x'^2-(y_a-y'_a)^2-z^2, 2x (y_a-y'_a) \right),\\
&&X'_{\mu}=\frac{1}{S\bar{S}}\left(-2x' z, x'^2-x^2-(y_a-y'_a)^2-z^2, -2x' (y_a-y'_a) \right).\label{bXi}
\end{eqnarray} 
Notice that the first term of (\ref{bulkH}) is just the solution without boundary \cite{Liu:1998bu} and the second term comes from the boundary effects. 

According to \cite{Liu:1998bu}, the on-shell quadratic action for $H_{ij}$ is given by
\begin{eqnarray}\label{I2}
I_2=\int_{M} dx^d z^{1-d} \left(\frac{1}{4} H_{ij}\partial_z H_{ij} -\frac{1}{2}H_{ij}\partial_j H_{zi}\right).
\end{eqnarray} 
Note that the terms on $Q$ do not contribute to the quadratic action. That is because
\begin{eqnarray}\label{I2Q}
(\delta I)_Q=\int_{Q} dx^d (K^{ij}-K \gamma^{ij})\delta \gamma_{ij}=0,
\end{eqnarray} 
which vanishes due to BCs (\ref{NBC}). Substituting (\ref{bulkH}) into (\ref{I2}), we derive
\begin{eqnarray}\label{I2HH}
I_2=\frac{1}{4}\frac{\Gamma[d+2]}{\pi^{d/2}\Gamma[d/2](d-1)}\int dx^d dx'^d \hat{H}_{ij}({\bf{x}})\Big{[}  \frac{\mathcal{I}_{ij,kl}}{s^{2d}} + \frac{\mathcal{\bar{I}}_{ij,kl}}{\bar{s}^{2d}} \Big{]}\hat{H}_{kl}({\bf{x'}}),
\end{eqnarray}
where 
\begin{eqnarray}
&&s^2=(x-x')^2+(y_a-y'_a)^2, \label{s} \\
&& \bar{s}^2=(x+x')^2+(y_a-y'_a)^2, \label{bs}\\
&&\mathcal{I}_{ij,kl}=\lim_{z\to 0}\frac{1}{2}\left( J_{ik} J_{jl} +J_{il} J_{jk}\right)-\frac{1}{d}\delta_{ij}\delta_{kl},\label{Iijkl}\\
&& \mathcal{\bar{I}}_{ij,kl}=\lim_{z\to 0}\frac{1}{2}\left( \bar{J}_{ik} \bar{J}_{jl} +\bar{J}_{il} \bar{J}_{jk}\right)-\frac{1}{d}\delta_{ij}\delta_{kl}. \label{bIijkl}
\end{eqnarray} 
From (\ref{I2HH}), we finally obtain the two point function of stress tensor for holographic BCFT
\begin{eqnarray}\label{TTfromHH}
<T_{ij}({\bf{x}})T_{kl}({\bf{x'}})>=C_T\Big[ \frac{\mathcal{I}_{ij,kl}}{s^{2d}} +  \frac{\mathcal{\bar{I}}_{ij,kl}}{\bar{s}^{2d}}  \Big],
\end{eqnarray}
where $C_T=\frac{2\Gamma[d+2]}{\pi^{d/2}\Gamma[d/2](d-1)}$.  
Note that the first term of (\ref{TTfromHH}) is just the two point function without boundary, and the second term of (\ref{TTfromHH}) is due to the boundary effect, which depends on boundary conditions. After some calculations, we rewrite (\ref{TTfromHH}) into the form used in  \cite{McAvity:1993ue},
\begin{eqnarray}\label{TTfromHHnew}
<T_{ij}({\bf{x}})T_{kl}({\bf{x'}})>&=&\frac{1}{s^{2d}} \big[ \delta(v) \delta_{ij}\delta_{kl}+\epsilon(v) (I_{ik}I_{jl}+I_{il}I_{jk})+(\beta(v)-\delta(v))(\hat{X}_{i}\hat{X}_{j}\delta_{kl}+\hat{X}'_{k}\hat{X}'_{l}\delta_{ij})\nonumber\\
&&\ \ \ \ \ \  -\left(\gamma(v)+\epsilon(v)\right)(\hat{X}_{i}\hat{X}'_{k} I_{jl}+\hat{X}_{j}\hat{X}'_{l}I_{ik}+\hat{X}_{i}\hat{X}'_{l}I_{jk} +\hat{X}_{j}\hat{X}'_{k}I_{il}  ) \nonumber\\
&&+\left(\alpha(v)-2\beta(v)+4\gamma(v)+\delta(v)+2\epsilon(v) \right) \hat{X}_{i}\hat{X}_{j}\hat{X}'_{k}\hat{X}'_{l} \big],
\end{eqnarray}
where 
\begin{eqnarray}\label{Osbornformula}
\begin{split}
&v=\frac{s}{\bar{s}}=\sqrt{\frac{(x-x')^2+(y_a-y'_a)^2}{(x+x')^2+(y_a-y'_a)^2}}, \\
&I_{ij}=\lim_{z\to 0} J_{ij}=\delta_{ij}-2\frac{(x_i-x'_i)(x_j-x'_j)}{s^2}, \\
&\hat{X}_i=\lim_{z\to 0} X_{i}=\frac{1}{s\bar{s}}\left(x^2-x'^2-(y_a-y'_a)^2, 2x (y_a-y'_a) \right) ,\\
&\hat{X}'_i=\lim_{z\to 0} X'_{i}=\frac{1}{s\bar{s}}\left(x'^2-x^2-(y_a-y'_a)^2, -2x' (y_a-y'_a) \right),
\end{split}
\end{eqnarray} 
and 
\begin{eqnarray}\label{abcde}
\begin{split}
&\alpha(v)=C_T \frac{(d-1) \left(1+ v^{2 d}\right)}{d}, \\
&\beta(v)=\delta(v)=-C_T\frac{1+ v^{2 d}}{d}, \\
&\gamma(v)=C_T \frac{ v^{2 d}-1}{2} ,\\
&\epsilon(v)=C_T \frac{1+ v^{2 d}}{2}.
\end{split}
\end{eqnarray} 
Note that $v$ (\ref{Osbornformula}) characterizes the distance to the boundary. In the limit far aways from the boundary we have $v=0$, while in the limit near the boundary we have $v=1$. 
It is remarkable that, the above functions take exactly the same form as those for free fermions and free scalars (half NBC and half DBC) \cite{McAvity:1993ue}. According to \cite{Alishahiha:2011rg}, this is the expected result and can be regarded as a check of our calculations.  Reflection positivity in Euclidean signature impose bounds on the functions (\ref{abcde}). According to \cite{Herzog:2017xha}, we have
\begin{eqnarray}\label{bounds}
\alpha(v)\ge 0,\  -\gamma(v)\ge 0, \ \epsilon(v)\ge 0 .
\end{eqnarray}
It is remarkable that our holographic results (\ref{abcde}) indeed satisfy the above positivity constraints. This is another support of our results. 

Now let us focus on the normal components of (\ref{TTfromHH}), from which we derive the Zamolodchikov norm of displacement operator
\begin{eqnarray}\label{CDfromHH}
C_D =\alpha(1)=\frac{4\Gamma[d+2]}{\pi^{d/2}d\Gamma[d/2]},
\end{eqnarray}
for $T=\rho=0$. Under the same conditions, the holographic Casimir coefficients (\ref{aN}) reduce to
\begin{eqnarray}\label{afromHH}
\alpha =-\frac{d \Gamma \left(\frac{d-1}{2}\right)}{\sqrt{\pi } \Gamma \left(\frac{d+2}{2}\right)}.
\end{eqnarray}
Comparing (\ref{CDfromHH}) with (\ref{afromHH}), we verify the universal relation (\ref{universalrelation2}) between displacement operator and Casimir effect. It is interesting to generalize the above discussions to $T\ne 0$. However, this is a non-trivial problem due to complicated BCs and we leave it to future work.


\section{Conclusions and Discussions}

In this paper, we have obtained universal relations between Casimir effect, Weyl anomaly and displacement operator for BCFTs in general dimensions. We verify our results by free scalars and holographic BCFTs. It is interesting to generalize our work to defect CFTs \cite{Billo:2016cpy}. Notice that BCFT can be regarded as a defect CFT with co-dimension one. And the case of co-dimension two is closely related to R\'enyi entanglement entropy and is studied by \cite{Bianchi:2016xvf,Bianchi:2015liz,Lewkowycz:2014jia,Dong:2016wcf,Chu:2016tps,Balakrishnan:2016ttg}. It is also interesting to derive the holographic two point functions of stress tensor and current for general boundary conditions. We hope we could address these problems in future. 

\section*{Acknowledgements}
We would like to thank M. Billò, V. Goncalves, E. Lauria, M. Meineri, M. Fujita, Y. Sato and Xinan Zhou for correspondence and useful comments. This work is supported by the funding of Sun Yat-Sen University.

\appendix

\section{Useful formulas}

In this appendix, we apply the method of \cite{Bianchi:2016xvf} to re-express a function as distributions. 
Let us start with the function 
\begin{eqnarray}\label{kernel}
K(x,y)=\frac{y^{2\a}}{(x^2+y^2)^{d+\b}}
\end{eqnarray} 
and a test function $f(y)$ which is regular at $y=0$ and decays fast enough when $y\to \infty$. Define the integral
\begin{eqnarray}\label{Ix}
I(x)=\int dy^{d-1} K(x,y) f(y).
\end{eqnarray} 
Since we are interested in the singular parts, we focus on the domain $|y|\le 1$. (\ref{Ix}) becomes
\begin{eqnarray}\label{Ix1}
I(x)=\sum_{n=0}^{\infty} \frac{1}{n!}\partial_{i_1}...\partial_{i_n}f(0)\int_{|y|\le 1} dy^{d-1} y^{i_1}...y^{i_n}K(x,y)+ \text{regular terms}.
\end{eqnarray}
Performing the coordinate transformation $y^a=x z^a$, we get
\begin{eqnarray}\label{Ix2}
I(x)&=&\sum_{n=0}^{\infty} \frac{1}{n!}\partial_{i_1}...\partial_{i_n}f(0)\frac{1}{x^{d-n+1+2\b-2\a}}\int_{|z|\le 1/x} dz^{d-1} \frac{z^{i_1}...z^{i_n}z^{2\a}}{(1+z^2)^{d+\b}}+ \text{regular terms}\nonumber\\
&=& \sum_{n=0}^{\infty} \frac{1}{n!}\partial_{i_1}...\partial_{i_n}f(0)\frac{\Omega_{d-2}}{x^{d-n+1+2\b-2\a}}\frac{\delta^{i_1i_2}...\delta^{i_{n-1}i_n}+\text{permutations}}{\text{normalization}} \times\nonumber\\
&&\ \ \ \ \ \ \ \ \int_{r\le 1/x} dr\frac{r^{d-2+n+2\a}}{(1+r^2)^{d+\b}}+ \text{regular terms}\nonumber\\
&=& \sum_{n=0}^{\infty} \frac{1}{n!}\partial_{i_1}...\partial_{i_n}f(0)\frac{1}{x^{d-n+1+2\b-2\a}}\frac{\delta^{i_1i_2}...\delta^{i_{n-1}i_n}+\text{permutations}}{\text{normalization}} \times\nonumber\\
&&\frac{\pi^{\frac{d-1}{2}}\Gamma \left(\a+\frac{d+n-1}{2} \right) \Gamma \left(\b-\a+\frac{d-n+1}{2}\right)}{\Gamma(\frac{d-1}{2}) \Gamma (\b+d)}+ \text{regular terms}\nonumber\\
\end{eqnarray}
In the weak limit, we can replace $f(0)$ by $\delta^{d-1}(y)$ and obtain
\begin{eqnarray}\label{Ix3}
K(x,y)=\frac{\pi^{\frac{d-1}{2}}}{\Gamma(\frac{d-1}{2}) \Gamma (\b+d)}
\sum_{n=0}^{\infty}P_n\frac{\Gamma \left(\a+\frac{d+n-1}{2} \right) \Gamma \left(\b-\a+\frac{d-n+1}{2}\right)}{x^{d-n+1+2\b-2\a}}
\end{eqnarray}
where
\begin{eqnarray}\label{Pn}
P_n=\frac{\partial_{i_1}...\partial_{i_n}\delta^{d-1}(y)}{n! }\frac{\delta^{i_1i_2}...\delta^{i_{n-1}i_n}+\text{permutations}}{\text{normalization}} .
\end{eqnarray}

From (\ref{Ix3}) and the derivatives of (\ref{Ix3}), we obtain the following useful formulas.
\begin{eqnarray}\label{keyformula}
&&\frac{y^ay^b}{(x^2+y^2)^{d+2}}=\frac{\pi^{\frac{d-1}{2}}\Gamma(\frac{d+1}{2})} {4\Gamma(d+2)} \left[ \delta^{ab}(\frac{(d+1)\delta^{d-1}(y)}{x^{d+3}}+\frac{\partial^2\delta^{d-1}(y)}{2x^{d+1}}) + \frac{\partial^a\partial^b\delta^{d-1}(y)}{x^{d+1}} \right]\\
&&\frac{1}{(x^2+y^2)^{d}}=\frac{\pi^{\frac{d-1}{2}}\Gamma(\frac{d+1}{2})} {\Gamma(d)} \left[ \frac{\delta^{d-1}(y)}{x^{d+1}} + \frac{\partial^2\delta^{d-1}(y)}{2(d-1)x^{d-1}} \right]+...\\
&&\frac{y^a}{(x^2+y^2)^{d+1}}=-\frac{\pi^{\frac{d-1}{2}}\Gamma(\frac{d+1}{2})} {2\Gamma(d+1)} \left[ \frac{\partial^a\delta^{d-1}(y)}{x^{d+1}}+ \frac{\partial^a\partial^2\delta^{d-1}(y)}{2(d-1)x^{d-1}} \right]+...\\
&&\frac{y^a}{(x^2+y^2)^{d+2}}=-\frac{\pi^{\frac{d-1}{2}}\Gamma(\frac{d+1}{2})} {4\Gamma(d+2)} \left[ \frac{(d+1)\partial^a\delta^{d-1}(y)}{x^{d+3}}+ \frac{\partial^a\partial^2\delta^{d-1}(y)}{2x^{d+1}} \right]+...\\
&&\frac{y^2}{(x^2+y^2)^{d+2}}=\frac{\pi^{\frac{d-1}{2}}\Gamma(\frac{d+1}{2})} {4\Gamma(d+1)} \left[ \frac{(d-1)\delta^{d-1}(y)}{x^{d+1}} + \frac{(d+3)\partial^2\delta^{d-1}(y)}{2(d-1)x^{d-1}} \right]+...
\end{eqnarray}
Using the above formulas, we can derive (\ref{TDW1},\ref{TDW2},\ref{TDW3}).

\end{document}